# Finite State Machine based Vending Machine Controller with Auto-Billing Features

Ana Monga[1], Balwinder Singh[2]

[1,2]Academic and Consultancy Services-Division,
Centre for Development of Advanced Computing(C-DAC), Mohali, India
`er.ana27@gmail.com,balwinder.cdacmohali@gmail.com`

## ABSTRACT

*Nowadays, Vending Machines are well known among Japan, Malaysia and Singapore. The quantity of machines in these countries is on the top worldwide. This is due to the modern lifestyles which require fast food processing with high quality. This paper describes the designing of multi select machine using Finite State Machine Model with Auto-Billing Features. Finite State Machine (FSM) modelling is the most crucial part in developing proposed model as this reduces the hardware. In this paper the process of four state (user Selection, Waiting for money insertion, product delivery and servicing) has been modelled using MEALY Machine Model. The proposed model is tested using Spartan 3 development board and its performance is compared with CMOS based machine.*

## 1. INTRODUCTION

Vending Machines are used to dispense various products like Coffee, Snacks, and Cold Drink etc. when money is inserted into it*.* Vending Machines have been in existence since 1880s. The first commercial coin operated machine was introduced in London and England used for selling post cards. The vending machines are more accessible and practical than the convention purchasing method. Nowadays, these can be found everywhere like at railway stations selling train tickets, in schools and offices vending drinks and snacks , in banks as ATM machine and provides even diamonds and platinum jewellers to customers. Previous CMOS and SED based machines are more time consuming than the FPGA based machines [7]. The FPGA based machine is also more flexible, programmable and can be re-programmed. But in microcontroller based machine, if one wants to enhance the design, he has to change the whole architecture again but in FPGA user can easily increase the number of products.

In this paper a new approach is proposed to design an FSM based Vending Machine [3] with auto-billing features. The machine also supports a cancel feature means that the person can withdraw the request and the money will be returned back to the user. The user will get a bill of total number of products delivered with total price. This machine can be used at various places like Hotels, Restaurants and food streets. This reduces the time and cost.

International Journal of VLSI design & Communication Systems (VLSICS) Vol.3, No.2, April 2012

## 1.1 Operation of Vending Machine

I. When the user puts in money, money counter tells the control unit, the amount of money inserted in the Vending Machine.
II. When the user presses the button to purchase the item that he wants, the control unit turns on the motor and dispenses the product if correct amount is inserted.
III. If there is any change, machine will return it to the user.
IV. The machine will demand for servicing when the products are not available inside the machine.

## 1.2 FSM (Finite State Machine) [2] [3]

In a Finite State Machine the circuit's output is defined in a different set of states i.e. each output is a state. A State Register to hold the state of the machine and a next state logic to decode the next state. An output register defines the output of the machine. In FSM based machines the hardware gets reduced as in this the whole algorithm can be explained in one process.

Two types of State machines are:

**MEALY Machine**: In this machine model, the output depends on the present state as well as on the input. The MEALY machine model is shown in figure 1.

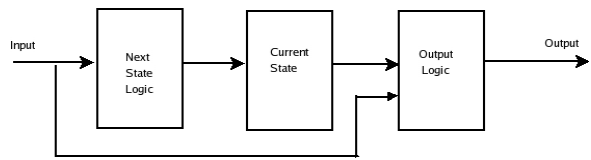

Figure 1: MEALY Machine Model

**MOORE Machine**: In Moore machine model the output only depends on the present state. The MOORE machine model is shown in figure 2.

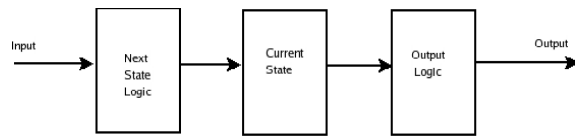

Figure 2: MOORE Machine Model

The paper is organized as: section 2 describes the related work. Section 3 relates the implementation of Vending Machine and section 4 gives the design methodology with description of states. Section 5 and section 6 shows simulation results and conclusion respectively.

## 2. RELATED WORK

Various researches have been carried out in order to design the Vending Machines. A few of them are discussed here as: Fauziah Zainuddin [1] proposes a vending machine for steaming frozen food using conceptual modelling. In which the process of three main states (user selection state, freezer state and steaming state) has been modelled using process approach, which emphasized on the process flow or control logic to construct the model for steamed buns vending machine application. Conceptual modelling is described in [6]. In [4] the concept of automatic mobile payment is discussed. This concept is based on the short message payment with the main control module M68HC11 and GPRS module MC35. Thee various methods of designing VHDL based machines are discussed in [2], [3] and [9]. Also in [5] the passenger's requirements for ticketing system are given. In [7] a coffee vending machine is designed using single electron encoded logic (SEEL). The designed circuit is tested and its power and switching time is compared with the CMOS technology.

## 3. IMPLEMENTATION OF VENDING MACHINE

In this paper a state diagram is constructed for the proposed machine which can vend four products that is coffee, cold drink, candies and snacks. Four select (select1, select2, select3, select4) inputs are taken for selection of products. Select1 is used for the selection of snacks. Similarly select2, select3, select4 are used for coffee, cold drink and candies respectively. Rs_10 and rs_20 inputs represents rupees 10/- and 20/- notes respectively. A cancel input is also used when the user wants to withdraw his request and also the money will be returned through the return output. Return, product and change are the outputs. Return and change vectors are seven bits wide. Money is an in/out signal which can be updated with the total money of all products delivered at a time. Money signal is seven bits wide. Money_count is an internal signal which can be updated at every transition. This signal is also seven bits wide. If the inserted money is more than the total money of products then the change will be returned through the change output signal. The products with their prices are shown by table 1. There are also two input signal clk and reset. The machine will work on the positive edge of clock and will return to its initial state when reset button is pressed. The proposed vending machine is designed using FSM modelling and is coded in VHDL language. The detail of the entire signal with their direction and description is shown in table 2.

Table 1: Products with their prices

| S.No. | Products | Price |
|---|---|---|
| 1. | Snacks | 30/- |
| 2. | Coffee | 40/- |
| 3. | Cold drink | 40/- |
| 4. | Candies | 30/- |

Table 2: Inputs/Outputs with Remarks

| Name | Width | Direction | Description |
|---|---|---|---|
| clk | 1 | input | Clock |
| Reset | 1 | Input | Syn reset |
| Sel1 | 1 | input | Snacks |
| Sel2 | 1 | input | Coffee |
| Sel3 | 1 | input | Cold drink |
| Sel4 | 1 | input | Candies |
| Cancel | 1 | input | Cancel |
| Money | 7 | inout | Total money |
| Rs_10 | 1 | input | Rupees 10/- |
| Rs_20 | 1 | input | Rupees 20/- |
| Product | 1 | output | Product out |
| Change | 7 | output | Extra change |
| Return | 7 | output | Return money |

## 4. DESIGN METHODOLOGY

The state diagram mainly consists of four states (User Selection, Waiting for the money insertion, product delivery and servicing (when product_not_available='1')). Initially when the reset button is pressed, the machine will be ready for the users to select the product. This state is the initial state of the design. After this the user will select the product to be dispensed. This state can be one of the select1, select2, select3 and select 4. The machine can accept only two types of notes i.e. rupees 10/- and 20/-. Let us suppose that the user selects sel1 input. The machine will firstly check that whether the products are available in the machine or not. After this the control unit will move to the waiting state, where it will wait for the money to be inserted. Then if rupees 10/- note is inserted then the machine will go to state_1 and wait until the desired money is inserted. And if rupees 20/- note is inserted the machine will move to state_2 and then wait until 30/- rupees are inserted to the machine. When the desired amount is inserted the machine will go to the snacks state and snacks will be delivered at the product output. If products are not available in the machine then the control unit will demand for servicing and after service the machine will get reset. This methodology is explained using a flow diagram shown in figure 3.

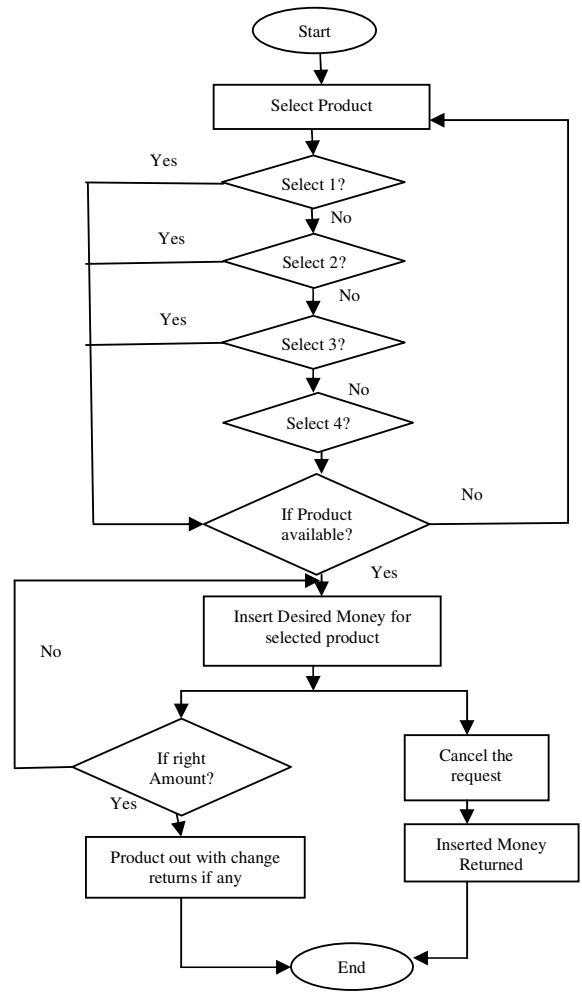

Figure 3: Flow Chart for Proposed Vending Machine

There is also an additional feature of withdrawing the request if the user doesn't want to take the product. When cancel button is pressed then the money inserted will be returned to the user through the return output. A money_count signal is used for calculating the total money inserted in the machine. And if the money inserted is more than the money of the product then the extra change will be returned to the user. The total amount of the product taken at a time is shown by the money signal. Similarly the user can select and get the other products following the above procedure.

**Description of states**

The selection of products and all the states are shown below in figure 4.

- When initialize=>
    - money_count=0;
    - Change=0;
    - Product=0;
- When select1=>
    - Sel1&!sel2&!sel3&!sel4
    - When product_available=1 => nx_st1<= waiting1;
    - When product_available=0 => nx_st1<= service1;
- When waiting1=>
    - When rs_10&!rs_20 => nx_st1<=state_1;
    - When !rs_10&rs_20=> nx_st1<= state_2;
    - Change=0;    product=0;
    - When money_count>=30  nx_st1<= snacks;
- When state_1=>
    - Rs_10=1 & rs_20=0;
    - Change=0;    Product=0;
    - Money_count=money_count+10;

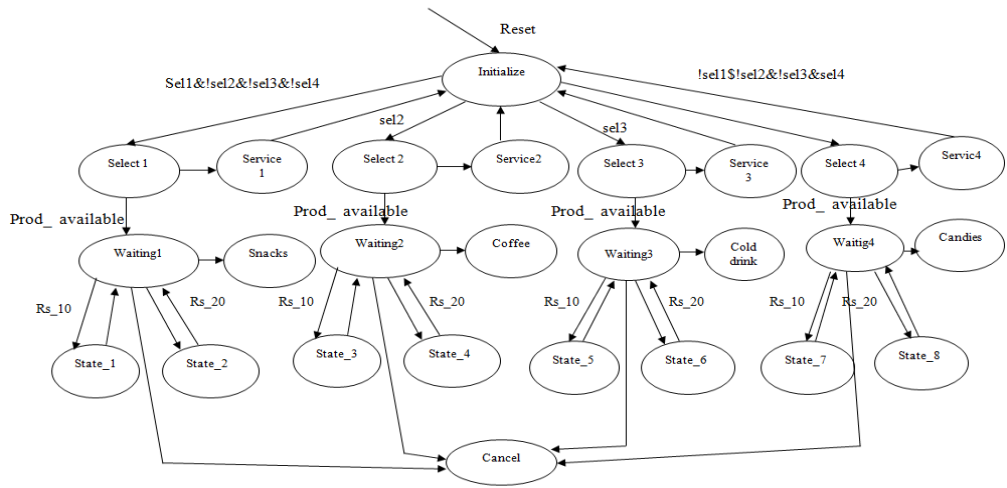

Figure 4: Finite State Machine Diagram of Vending Machine

- When state_2=>
    - Rs_10=0 & rs_20=1;
    - Change=0;        Product=1;
    - Money_count=money_count+20;
- When snacks=>
    - Money_count>=30;
    - Product=1;
    - Change=money_count-30;
    - Snack_count=snack_count-1;
- When service=>
    - snack_count= 4
    - product<=0;
    - next_state<=resett;
- When cancel1=>
    - cancel=1;
    - return<=money_count;

Similarly we can select other products (coffee, Cold drink and candies).

## 5. SIMULATION RESULTS

The state diagram shown in figure 4 is simulated using Xilinx ISE Simulator. Simulation Waveforms for the selection of four products like snacks is shown in figure 5 and 6 respectively with servicing feature when products are not available in the machine and change return features when the money inserted is more than the money of the product.

Let us take an example that the user wants to take Snacks. When one selects sel1 button, the machine will check that whether the products are available or not, if available then it will go to the waiting state and wait for total money insertion. If rs_10 note is inserted it will go to state_1 and if rs_20 note is inserted it will so to state_2 and check whether money_count>=30 or not. If the money_count >- 30 then machine will go to state snacks and vend the product.

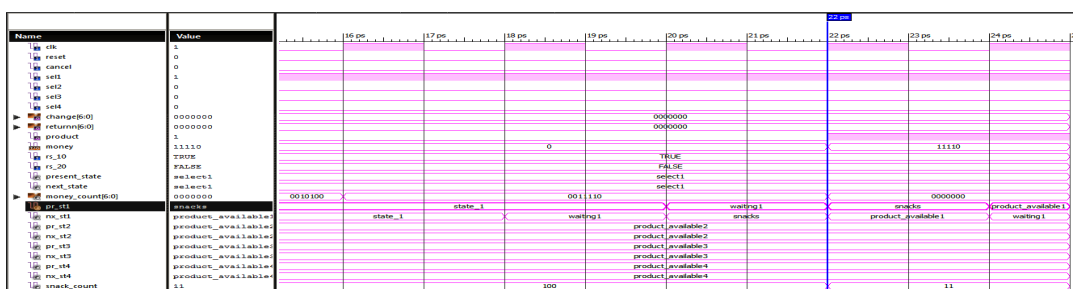

Figure 5: Simulation waveform showing selection of Snacks

International Journal of VLSI design & Communication Systems (VLSICS) Vol.3, No.2, April 2012

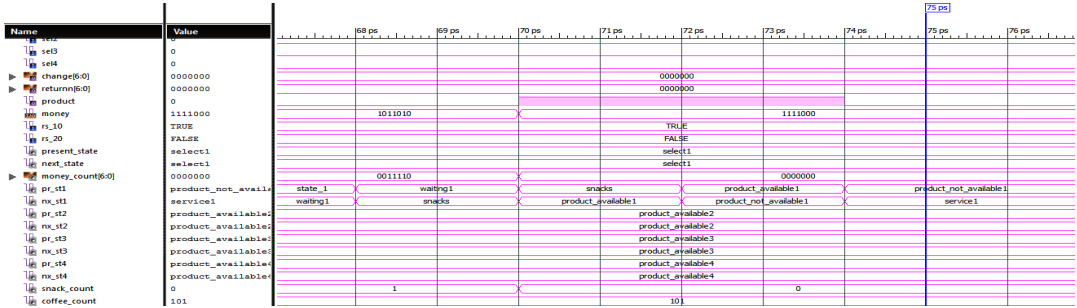

Figure 6: Simulation waveform when snack_count=0

If the user wants to cancel the request, can do so by pressing the cancel button and the whole money entered will be returned to the user. This is shown in figure 7.

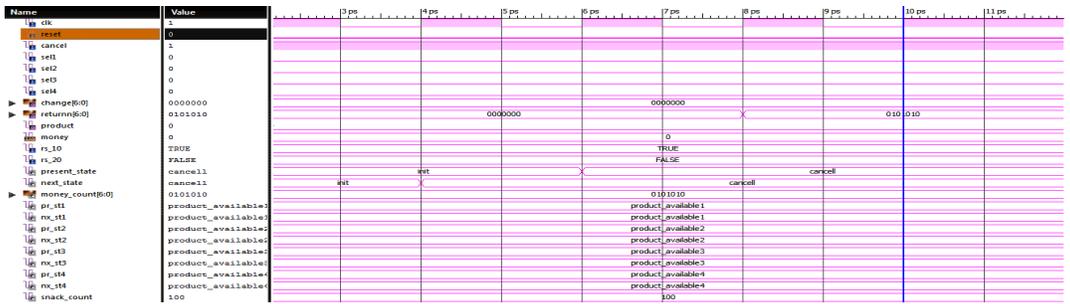

Figure 7: Simulation waveform showing Cancel Operation

The RTL view of the machine is shown in figure 8.

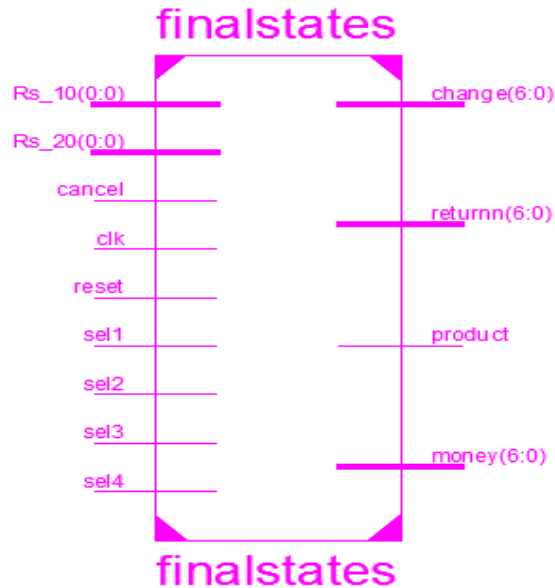

Figure 8: RTL view of Vending Machine.

The design summary of the proposed machine is shown below in table 3.

Table 3: Device Utilization Summary

| Logic utilization | Used | Available | Utilization |
|---|---|---|---|
| Number of Slices | 228 | 3584 | 6% |
| Number of Slice flip flops | 98 | 7168 | 1% |
| Number of 4 input LUT's | 432 | 7168 | 6% |
| Number of bonded IOB's | 31 | 173 | 17% |
| Number of GCLK's | 1 | 8 | 12% |

The comparison of VHDL based machine with the CMOS technology based machine on the basis of switching speed is shown in table 4.

Table 4: Comparison of Switching Speed

| Parameter | using FPGA | Using Single Electron Encoded Logic (SEEL) | Using CMOS technology |
|---|---|---|---|
| Switching Speed | 9.3 ns | 12.53 ns | 300ns |

## 6. CONCLUSION

The present FPGA based vending machine controller is implemented using FSMs with the help of Xilinx ISE Design Suite 12.4. The design is verified on the FPGA Spartan 3 development Board. State machines based vending Systems enhances productivity, reduces system development cost, and accelerates time to market. Also FPGA based vending machine give fast response and easy to use by an ordinary person. The designed machine can be used for many applications and we can easily enhance the number of selections. The next stage of this study is to convert this model into hardware and to calculate the total power consumption of the machine.

International Journal of VLSI design & Communication Systems (VLSICS) Vol.3, No.2, April 2012

International Journal of VLSI design & Communication Systems (VLSICS) Vol.3, No.2, April 2012

## Authors Biography

**Balwinder Singh** has obtained his Bachelor of Technology degree from National Institute of Technology, Jalandhar and Master of Technology degree from University Centre for Inst. & Microelectronics (UCIM), Punjab University Chandigarh in 2002 and 2004 respectively. He is currently serving as Senior Engineer in Centre for Development of Advanced Computing (CDAC), Mohali and is a part of the teaching faculty and also pursuing Ph.D. from GNDU Amritsar. He has 6+ years of teaching experience to both undergraduate and postgraduate students. Singh has published three books and many papers in the International & National Journal and Conferences. His current interest includes Genetic algorithms, Low Power techniques, VLSI Design & Testing, and System on Chip.

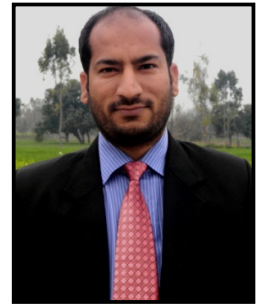

**Ana Monga** received the B.Tech. (Electronics and Communication Engineering) degree from the D.A.V Institute of Engineering and Technology, Jalandhar affiliated to Punjab Technical University, Jalandhar in 2006, and presently she is doing M.Tech (VLSI design) degree from Centre for Development of Advanced Computing (CDAC), Mohali and working on her thesis work. Her area of interest is FPGA Implementation and VLSI Design and Testing, Electronic devices and circuits, signals and systems.

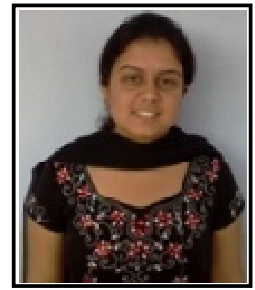